\documentclass[amsmath,showpacs]{revtex4}

\usepackage{graphicx}

\newcommand{\be}{\begin{equation}}
\newcommand{\ee}{\end{equation}}
\newcommand{\bea}{\begin{eqnarray}}
\newcommand{\eea}{\end{eqnarray}}
\newcommand{\ket}[1]{ |  #1  \rangle}
\newcommand{\bra}[1]{ \langle #1   |}

\newcommand{\mat}[1]{\begin{bmatrix}#1\end{bmatrix}}

\newcommand{\one}{\mbox{$1 \hspace{-1.0mm}  {\bf l}$}}

\newcommand{\tr}{\text{tr}}
\newcommand{\T}{^T}
\renewcommand{\d}{\text d}
\newcommand\ba{\begin{eqnarray}}
\newcommand\ea{\end{eqnarray}}
\newcommand{\mya}{\mathbf{a}}
\newcommand{\myb}{\mathbf{w}}
\newcommand{\mygcd}[2]{\gcd(#1,#2)}

\begin{document}

\title{Finite size effects in entangled rings of qubits}

\author{T. Meyer}
\email{tim.meyer@itp.uni-hannover.de}
\author{U. V. Poulsen}
\author{K. Eckert}
\author{M. Lewenstein}
\author{D. Bru\ss }

\affiliation{
Institut f\"ur Theoretische Physik, Universit\"at Hannover, 30167
Hannover, Germany
}


\begin{abstract}
We study  translationally invariant rings of qubits with a finite 
number  of sites $N$, and find the maximal nearest-neighbor entanglement
for a fixed $z$ component of the total spin.
For small numbers of sites our results are analytical. The use of  a
linearized version of the concurrence allows us to
relate the maximal concurrence to the ground state energy of an XXZ spin
model, and to calculate it numerically 
 for $N\leq 24$. We point out some interesting
finite-size effects. Finally, we generalize our results 
beyond nearest neighbors.
\end{abstract}



\maketitle


\section{Introduction}
\label{intro}

Entanglement in many-body systems such as spin chains
 has recently attracted much
attention. The reasons are at least twofold: on one side one is
interested in  understanding fundamental structures of 
entangled multipartite states  
and the natural occurrence of such states in systems
with spin-spin interactions, with quantum information processing
purposes in mind.
 On the other side insights from quantum information theory
may lead to a deeper understanding of many-body phenomena
in strongly interacting systems, such as quantum phase transitions
and critical behavior.

Much effort has been devoted to studying the natural appearance of
bipartite
entanglement in ground states of 1-dimensional spin models 
with an Ising
Hamiltonian or a Heisenberg-type 
Hamiltonian~\cite{fazio,nielsen,zanardi,vidal},
with special emphasis on properties of entanglement near a quantum
phase transition. It has  been pointed out that some entanglement
remains for finite temperatures (so-called thermal entanglement
\cite{vedral}), which for some transverse field can even increase
with increasing temperature. Recently, it has been shown that 
 maximally entangled states of atoms can be created 
by crossing a quantum phase transition \cite{dorner}.

A complementary approach to the phenomenon of entanglement in chains of
qubits has been initiated by O'Connor and Wootters 
~\cite{wootters} (see also~\cite{wchain}): they ask
(without specifying a Hamiltonian) the fundamental question ``what is
the maximal entanglement between two neighboring sites of an
entangled ring with translational invariance''?  An ``entangled ring"
is defined as a chain of qubits with periodic boundary conditions.
Due to the so-called ``monogamy of entanglement" it is impossible for
a site to be maximally entangled with both its neighbors: shared
entanglement is always less than maximal \cite{brusspra,oldwootters}.
In Ref.~\cite{wootters} the question of the upper limit for the nearest
neighbor entanglement was simplified by introducing two additional
restrictions on the allowed states: 
\begin{itemize}
\item[(i)] 
the state is an eigenstate of the $z$ component of the total
spin,
\item[(ii)] neighboring spins cannot both be ``up''.
\end{itemize}
  Both
restrictions are based on an educated guess for the optimal states for
the general problem.
 O'Connor and Wootters 
solved the restricted optimization problem 
by relating it to an effective Hamiltonian for the one-dimensional
ferromagnetic $XY$ model, and
 found the maximal nearest-neighbor concurrence to be
\begin{equation}
C^{\text{max}}(N,p)=\frac{2\sin\left(\frac{p\pi}{N-p}\right)}{N\sin\left(
    \frac\pi{N-p}\right)}
\label{eq:cwootters}\ ,
\end{equation}
where $N$ is the number of sites and $p$ is the number of
up-spins. 
Eq. \eqref{eq:cwootters}  provides a lower bound
for the problem without restriction (ii).
For given $N$ and $p$, it may or may not happen that $C$
can be increased by also allowing  states where two neighboring spins
are up.
In the limit $N\rightarrow\infty$, the optimal number of up-spins in
Eq.~\eqref{eq:cwootters} approaches $p_{opt}\approx
0.301\;N$. This leads to an  asymptotic
value of $C^\text{max}\approx 0.434$. Whether this number can be improved
by removing restriction (ii) or  restriction (i) is an open problem.
 We will
illustrate, however,  that for a fixed $p$
restriction (ii) tends to play a decreasing role as $N$ is increased.

In this article we mainly focus on the problem of optimizing the
nearest-neighbor
entanglement for given finite $N$ and $p$, {\em without}
imposing  restriction (ii).
Using the methods described below we
 solve a number of cases analytically. We supplement these results
by numerical calculations, and thus arrive at  a   rather complete
picture of the solutions for $N \le 24$. Some finite size effects that 
occur only for specific   small numbers of sites will be
pointed out.

The paper is organized as follows: In Sec.~\ref{maxent} we introduce
the problem and  our  notation, and study  symmetry under reflections. In
Sec.~\ref{sec:ana_res} we describe our analytical results 
for the maximal nearest-neighbor entanglement 
in the cases  $p=2$ and 
$p=N/2$. In Sec.~\ref{stability} we  numerically
investigate whether the solutions 
in \cite{wootters} are at least still \emph{local}
maxima when restriction (ii) is removed. In Sec.~\ref{sec:map} we
use a method recently introduced by Wolf {\it et al.}~\cite{cirac}, mapping
the original nonlinear problem onto a series of linear ones. We
discuss some general properties of the mapping, and  use it to
numerically calculate the maximal nearest-neighbor 
concurrence for  $N \le 24$. Finally, in
Sec.~\ref{sec:discuss} we discuss our results and point out some
interesting questions that are still open.

\section{Maximal nearest neighbor entanglement}
\label{maxent}

We consider a ring of $N$ qubits, out of which $p$ are in the state
$\ket{\uparrow}$ and $N-p$ are in the state $\ket{\downarrow}$.
In the following we will  use the notation $\ket{\downarrow}\equiv\ket{0}$
and $\ket{\uparrow}\equiv\ket{1}$, and will refer to 
$p$ as the occupation number. We will always consider
translationally invariant states.

\subsection{Translationally invariant states and concurrence}
\label{transstates}

Our aim is to calculate the maximal nearest-neighbor entanglement of
two qubits,
using the concurrence as a measure. 
The concurrence  is defined as~\cite{defconc}
$C(\rho)=\max\{\lambda_1-\lambda_2-\lambda_3-\lambda_4,0\}$, where
$\lambda_1\ge\lambda_2\ge\lambda_3\ge\lambda_4\ge0$
are the square roots of the eigenvalues of $\rho\tilde\rho$,
and 
$\tilde\rho:=(\sigma^y\otimes\sigma^y)\rho(\sigma^y\otimes\sigma^y)$
is the spin-flipped density matrix.

The structure of any two-qubit reduced density matrix
on the ring is, due to the properties of translational
invariance and the occupation number  $p$ being fixed, given by 
\begin{equation}
\label{eq:rho2}
\varrho^{(2)}=\mat{ v & 0 & 0& 0\\ 0 & w & z & 0 \\ 0 & z^* & w & 0 \\
  0 & 0 & 0 & y} 
\end{equation}
with $v+2w+y=1$. The basis is chosen as $(\ket{00},\ket{01},\ket{10},
\ket{11})$.  Here the zero entries are due to considering a
fixed $p$. The diagonal entries in
the density matrix are real, and 
 the
equality of two of the diagonal elements is caused by the translational
invariance.
  Note 
that $v=y+(N-2p)/N$, i.e.\ the number of
down-down  pairs of neighboring spins
depends on the number of up-up pairs.
We will see below that for states with maximal entanglement 
the off-diagonal element $z$
can be taken to be real.

The concurrence of the density matrix~\eqref{eq:rho2} is given
by~\cite{wootters} 
\begin{equation}C=2\max\{|z|-\sqrt{vy},0\}\ .
\label{concurrence}
\end{equation}
As one observes from Eq.~\eqref{concurrence},
 the task of finding the state that
maximizes the nearest-neighbor concurrence is substantially simplified by
requiring $y=0$, i.e.  two neighboring spins cannot be both ``up''.
This is the assumption that was made in \cite{wootters} 
---  we will relax this constraint
throughout our work. 

 For given $N$ and
$p$ we apparently need to maximize Eq.~(\ref{concurrence}) over all
translationally invariant states, that is, 
over all density matrices fulfilling
$[\varrho,\mathcal{T}]=0$, where $\mathcal{T}$ is the operator that
implements a translation  by one site. As was shown
in~\cite{wootters}, the problem splits into $N$ maximization
problems over \emph{pure} states: since $\varrho$ and $\mathcal{T}$
commute, the spectral decomposition of
 $\varrho$ can be expressed in projectors onto
eigenstates of $\mathcal{T}$, namely 
\begin{equation}
  \label{eq:decomp_rho}
  \varrho=\sum_{k=0}^{N-1} p_k \ket{\psi_k}\bra{\psi_k},
\end{equation}
where
$p_k\geq 0$ and $\sum_kp_k=1$. Here the eigenstates of the translation
operator are defined by
$\mathcal{T}\ket{\psi_k}=e^{i\frac{2\pi}{N}k}\ket{\psi_k}$. Since $C$
is a convex function, we have
\begin{equation}
C\left(\sum_{k} p_k \ket{\psi_k}\bra{\psi_k}\right)\le \sum_kp_k
C(\ket{\psi_k}\bra{\psi_k})
\le \max_k\{C(\ket{\psi_k}\bra{\psi_k})\}
,
\end{equation}
and it is thus sufficient to first maximize the concurrence over pure
states within each eigenspace of the translation operator, and then
choose the maximal result over all eigenstates.  In fact, the
situation is even simpler, since, as shown in~\cite{wootters}, the
search can be restricted to the $k=0$ eigenstates, i.e. to eigenstates
of $\mathcal{T}$ with eigenvalue 1.

For a given number of sites $N$ and a given occupation number $p$, the
most general eigenstate of $\mathcal{T}$ with eigenvalue 1 is a 
superposition that looks as follows: 
\begin{equation}
\ket{\psi(N,p)} = \sum_{\mu}a_\mu\ket\mu\ , 
\label{eq:genstate}
\ee
with 
the normalizations 
\be
\bra\mu\mu\rangle=1\quad\text{and}\quad\sum_\mu |a_\mu|^2=1\ ,
\label{norm}
\end{equation}
and with
\be
\ket\mu = \mathcal{N}_\mu\sum_{j=0}^{N-1}\mathcal T^j\ket{\phi_\mu}\ .
\label{mu}
\ee
The above notation is as follows: the index $\mu$ enumerates all
distinct translationally invariant configurations of the
 ring where the  $p$
up-spins and $(N-p)$ down-spins  have given {\em relative}
positions. The state
$\ket{\phi_\mu}$ is one representative member of the $\mu$th configuration.
The operator
inducing a translation by $j$ sites is given by $T^j$, and
$\mathcal{N}_\mu$ denotes an appropriate normalization factor.
Thus,  $\ket\mu$ is a normalized eigenstate of
$\mathcal{T}$ corresponding to the configuration $\mu$ and the eigenvalue
1. Note that for a given $\mu$ it may happen that
$\mathcal{T}^j\ket{\phi_\mu}=\ket{\phi_\mu}$ already for $j=\tilde{N}<N$; in
that case it is not possible to construct eigenstates corresponding to
all eigenvalues of $\mathcal T$ from $\ket{\phi_\mu}$. This is accounted for 
by an appropriate value of $\mathcal{N}_\mu$, 
which in the general case ($\tilde{N}\le N$) reads
$\mathcal{N}_\mu=\sqrt{\tilde{N}}/N$.

We use the following convention to denote $\mu$ explicitly: 
the composite index $\mu$
consists of $p$ entries, where each entry specifies a distance from one
up-spin to the next up-spin in $\ket{\phi_\mu}$. We choose the first
entry to be the smallest. Note that the $p$ entries add up to $N$. An
example will clarify this notation: let us consider the case of $N=5$
qubits and the occupation number $p=2$. Then the index $\mu$ can only
take two combinations of entries, namely 1,4 and 2,3.  The most
general translationally invariant state with translational eigenvalue
$1$ is written as \bea \ket{\psi(N=5,p=2)} &=&
a_{1,4}\ket{1,4}+a_{2,3}\ket{2,3}
\nonumber \\
&=& \sum_{\mu=1,4;2,3}a_\mu \frac{1}{\sqrt{5}}\sum_{j=0}^4
{\mathcal T}^j\ket{\phi_\mu}\nonumber \\
&=& a_{1,4}\frac{1}{\sqrt{5}} \left(
  \ket{11000}+\ket{01100}+\ket{00110}+\ket{00011} +\ket{10001}
\right)+  \nonumber \\
& & a_{2,3}\frac{1}{\sqrt{5}} \left(
  \ket{10100}+\ket{01010}+\ket{00101}+\ket{10010} +\ket{01001} \right)
\ ,
\label{N=5}
\eea
where the normalization reads $|a_{1,4}|^2+|a_{2,3}|^2=1$.
The task is then to find the optimal coefficients $a_\mu$, such that 
$C$ is maximized.

\subsection{Reflection symmetry}
\label{generalvzy}
One main big difficulty in the maximization problem of the
concurrence is the growing number of free parameters $a_\mu$ for 
higher $N$ and $p$~\footnote{
One can show that $\binom{N}{p}/N$ is a lower bound for the number of
different $\mu$'s.}.
In this subsection we will elaborate on a symmetry argument that
eliminates nearly half of the parameters.

Recall the state under consideration, Eq.~(\ref{eq:genstate}), and the
definition of the $p$-dimensional index $\mu$, which specifies the
number of steps from one up-spin to the next. It is clear that a
cyclic permutation of the entries of $\mu$ will always result in the
same state $\ket\mu$, whereas an anti-cyclic one will result in a
state where all spins are reflected around a certain site. Since
$\ket{\mu}$ is translationally invariant, it does not matter around
which site the reflection is performed, and for convenience we can
assume that the reflection interchanges the two sites that we are
focusing on.  Denoting the reflection operator by $\mathcal R$, it is
then obvious that $\varrho\rightarrow\mathcal{R}\varrho\mathcal{R}$
leads to $\varrho^{(2)}\rightarrow\bigl[\varrho^{(2)}\bigr]^\text{T}$
(in our chosen basis) for any translationally invariant $\varrho$. In
terms of the matrix elements, reflection causes $v,w,y\rightarrow
v,w,y$ while $z\rightarrow z^*$.  In particular the reflected
state has the same concurrence as the original one,
$C(\mathcal{R}\varrho\mathcal{R})=C(\varrho)$.

Now, let us study the entanglement properties of
$\tilde{\varrho}=(\varrho+\mathcal{R}\varrho\mathcal{R})/2$. We arrive
at
\begin{equation}
  \label{eq:c_rtwirl}
  C(\tilde{\varrho})
  =
 |z+z^*|-2\sqrt{\frac{1}{2}(v+v)\frac{1}{2}(y+y)}
  =C(\varrho)
  ,
\end{equation}
and thus it is sufficient to consider states of the form
$\tilde{\varrho}$. Note that $\tilde\varrho$ commutes with the
reflection operator, i.e.  $[\tilde\varrho,\mathcal{R}]=0$, and
therefore can be decomposed into projectors onto eigenstates of the
reflection operator.  By the same argument that allowed us to reduce
the original problem to a pure state problem for each translational
eigenvalue, we can now restrict our attention to pure states with
translational eigenvalue 1 \emph{and} odd or even reflection symmetry.
In fact, using the observation from \cite{wootters} that one only
needs to consider positive coefficients $a_\mu$ in the optimal state,
we immediately find that the even states will contain an optimal
state.  This result is non-trivial, intuitive as it may seem: A
non-linear optimization problem could lead to solutions that break the
underlying symmetries. Here, however, this is not the case and this
fact reduces the dimensionality of the relevant parameter space
significantly.

\section{Analytical results}
\label{sec:ana_res}

In the following we will consider various occupation numbers, starting
with the trivial case $p=1$: here the state is completely fixed
and reads
\be
\ket{\psi(N,p=1)}=\frac{1}{\sqrt{N}}\sum_{i=0}^{N-1}
{\mathcal T}^i\ket{\phi_{N}}\ ,
\ee
which is a generalized W-state. The concurrence 
between nearest neighbors (in fact, between any two sites)
is given as $C(N,p=1)=2/N$.  

\subsection{Maximal nearest neighbor concurrence for occupation number $p=2$}
\label{pequal2}

The case $p=2$ is already much more involved. The concurrence of a
given state can still be written in a compact way for all $N$, but it
will turn out that the optimization problem can only be solved
analytically for some small numbers of sites.  The most general
translationally invariant state for $N$ sites with occupation number
$p=2$ is given by 
\begin{equation}
 \ket{\psi(N,p=2)} 
 =
 \sum_{\mu}a_\mu\mathcal
 N_\mu\sum_{i=0}^{N-1} {\mathcal T}^i\ket{\phi_\mu}\ ,
 \label{p=2}
\end{equation}
where, as explained above, $\mu$ is a  index with two entries
which describes the
location of the up-spins in the ring; the first value for $\mu$ is $1,N-1$.
As mentioned above, for the optimization problem 
we can take the coefficients $a_\mu$ to be real and positive.

At this point we have to distinguish the cases of even and odd $N$:
\begin{itemize}
\item[(1)] {\em Even $N$:} \newline
The index $\mu$ stands for all inequivalent decompositions of
$N$ into a sum of two terms, starting at $1,N-1$, then $2,N-2$ and
so forth until $N/2,N/2$. The last term of the superposition
in Eq.~(\ref{p=2}) is special, because a translation by $N/2$ sites
already leads to 
$\mathcal{T}^{\frac{N}{2}}\ket{\phi_{N/2,N/2}}=\ket{\phi_{N/2,N/2}}$.
Therefore this contribution
 has a different 
normalization, as mentioned below Eq. (\ref{mu}).

This becomes immediately clear from looking at an example;
let us choose  $N=4$ and $p=2$:
\bea
\ket{\psi(N=4,p=2)}&=& a_{1,3}\ket{1,3}+a_{2,2}\ket{2,2}\nonumber \\
&=& \sum_{\mu=1,3;2,2}a_\mu\mathcal N_\mu\sum_{i=0}^{N-1}
{\mathcal T}^i\ket{\phi_\mu}\nonumber \\
&=&a_{1,3}\frac{1}{2}\left( \ket{1100}+\ket{0110}+\ket{0011}
+\ket{1001} \right)+  \nonumber \\
& & 
a_{2,2}\frac{1}{\sqrt{2}}\left( \ket{1010}+\ket{0101}\right)   \ .
\eea
For better clarity we will from now on
simplify the above notation of the composite index $\mu$ by dropping the second
entry, i.e. $a_{1,N-1}\equiv a_1$ and $\ket{1,N-1}\equiv \ket{1}$ and so forth. 
(This short-hand notation is of course only possible for the case $p=2$.)
Calculating the  matrix elements for even $N$, one finds
\bea
y&=&\frac1Na_1^2\label{eq:y}\ ,\\
v&=&\left\{\begin{array}{ll}\frac1N\left(N-4+a_1^2\right)& \text{for } \ \ 
N>3\\
0&\mbox{otherwise}\end{array}\right.\label{eq:v}\ ,\\
z&=&\frac2N\left(\sum\limits_{i=1}^{N/2-1}a_i a_{i+1}+\sqrt2a_{N/2-1}
a_{N/2}\right),\label{eq:z}
\eea
The concurrence is thus given by the expression
\be
C=\frac{2}{N}\, \max\left\{2|a_{1}a_{2}+a_{2}a_{3}+\dots
+a_{N/2-2}a_{N/2-1}+\sqrt{2}a_{N/2-1}a_{N/2}|
-|a_{1}|\sqrt{N-4+a_{1}^2},0\right\}\ .
\label{even2}
\ee
\item[(2)]{\em Odd $N$:} \newline
Here the double index $\mu$ runs from $1,N-1$ until $(N-1)/2,(N+1)/2$.
The
normalization of all terms in the superposition is identical.
An example of the most general state has already been shown 
for $N=5$ in Eq.~(\ref{N=5}).
With the short-hand notation introduced above 
we find
\bea
y&=&\frac1Na_1^2\ ,\\
v&=&\left\{\begin{array}{ll}\frac1N\left(N-4+
a_1^2\right)& \text{for } \ \ N>3\\
0&\mbox{otherwise}\end{array}\right.\ , \\
z&=&\frac2N\left(\sum\limits_{i=1}^{(N-1)/2-1}a_i a_{i+1}+
\frac12a_{(N-1)/2}^2\right)
\eea
The only differences to the case of even $N$ are the last term in $z$ being
$\frac12a_{(N-1)/2}^2$ instead of $\sqrt2a_{N/2-1}a_{N/2}$, and the
upper summation bounds for $z$.
 The concurrence for odd $N$ reads
\be
C=\frac{2}{N}\, \max\left\{2|a_{1}a_{2}+a_{2}a_{3}+\dots
+a_{(N-1)/2-1}a_{(N-1)/2}+\frac12 a_{(N-1)/2}^2|-|a_{1}|
\sqrt{N-4+a_{1}^2},0\right\}\ .
\label{odd}
\ee
\end{itemize}

From Eqs.~(\ref{even2}) and (\ref{odd}) we see immediately that the first
amplitude $a_{1}$ has a special role: it is the only one that appears
explicitly in the last term of the concurrence,
namely the term $-\sqrt{vy}$.
Note that the coefficient $a_1$ specifies the only case where two neighboring
spins are up.  Considering both terms of the concurrence where $a_1$ appears,
we can find a simple argument when the coefficient $a_1$ has to be zero in
order to maximize the concurrence: for $N\geq 8$, the inequality
\begin{equation}
2a_{2}-\sqrt{N-4+a_{1}^2}\leq 0
\end{equation}
holds for {\em any} $a_{2}$ and $a_{1}$.  Therefore, the concurrence
is always {\em increased} by setting $a_{1}=0$.  Thus, for more than 8
sites (for both even and odd $N$) and the occupation number $p=2$ one
finds indeed the highest concurrence by using the constraint ``no
neighboring spins are up'', i.e.\ $C^\text{max}$ is given by
Eq.~\eqref{eq:cwootters}. Note, however, that the actual coefficients
of the optimal state are not trivial to find from the method in
\cite{wootters}. Thus we have also used an alternative method which
employs Lagrange multipliers. Our method does not only provide the
value of the maximal concurrence, but also the coefficients $a_\mu$
that define the corresponding state. The details are described in the
Appendix~\ref{sec:lagrange}.  Using this method we find e.g. for the
case $N=8$ and $p=2$ the results $C^\text{max}(N=8,p=2)=\sqrt{3}/4$,
where the coefficients of the optimally entangled state are given by
$a_2=\sqrt{1/6}, a_3=\sqrt{1/2}$ and $a_4=\sqrt{1/3}$.

For $N<8$ we have in general $a_1\neq 0$, and the simple method 
from the Appendix cannot be
applied. One has to maximize the concurrence explicitly. This was done for the
cases $N=4$ (here it turns out that one can reach the same concurrence as for
$p=1$, by taking $a_1=\sqrt{1/3}$ and $a_2=\sqrt{2/3}$); for $N=5$ we have
only two terms, namely
$\ket{N=5,p=2}=a_1\ket{1,4}+a_2\ket{2,2}$
and arrive at the optimal solution $a_1=0.298$ and $a_2=0.955$.  The
case $N=6$ can again be solved analytically: if the state is written as
$\ket{N=6,p=2}=a_1\ket{1,5}+a_2\ket{2,4}+a_3\ket{3,3}$, the concurrence
reads 
\begin{equation}
C(N=6,p=2)=\frac{2}{3}\left(\sqrt{1-a_1^2-a_3^2}(a_1+\sqrt{2}a_3)-\frac12
a_1 \sqrt{2+a_1^2}\right)\ , 
\end{equation}
where we have eliminated $a_2$ via the
normalization constraint. The maximum can be found by demanding that
the derivatives with respect to $a_1$ and $a_3$ have to be zero. This
leads to the solution $a_1=0$ and $a_2=a_3=\sqrt{1/2}$ for the maximal
concurrence. Finally, for $N=7$ we have
$\ket{N=7,p=2}=a_1\ket{1,6}+a_2\ket{2,5}+a_3\ket{3,4}$ and 
\begin{equation}
C(N=7,p=2)
=\frac27\left(2a_2\sqrt{1-a_1^2-a_2^2}-(a_1-a_2)^2+1-a_1\sqrt{a_1^2+3}\right),
\end{equation}
where  this time we have applied the normalization to
eliminate $a_3$. Calculating $\d C/\d a_1$, we find that
$a_1>a_2\Rightarrow \d C/\d a_1<0$, whereas $\d C/\d a_2=0$ is
equivalent to $a_1-a_2=(a_2^2-a_3^2)/a_3$.  The last equality can only
be fulfilled for (I) $a_1>a_2>a_3$, where as noted $\d C/\d a < 0$,
for (II) $a_1=a_2=a_3$ which is clearly not optimal, or for (III)
$a_3>a_2>a_1$. For the last case, we can use again an argument like
for $N\ge 8$ to show that $a_1=0$, $a_2=\sqrt{(5-\sqrt5)/10}$, and
$a_3=\sqrt{(5+\sqrt5)/10}$ is optimal~\footnote{The boundaries $a_2=0,1$
are also easily seen to be non-optimal.} and gives
$C^\text{max}(N=7,p=2)=(1+\sqrt5)/7$.

This  completes the study of the case $p=2$. To summarize, we have found
a simple argument to show the solution 
in \cite{wootters} to be optimal for $N\ge 8$ and
$p=2$, and have
explicitly performed the optimization for rings with seven or less spins.  In
the case $N=5$, the seemingly simple one-parameter optimization results in a
fourth-order equation that had to be solved numerically.
All other cases were solved analytically. 
Note that $N=5$ is the only non-trivial case where the optimal state
contains two neighboring up-spins.
We summarize the results for
occupation number $p=2$ in Tab.~\ref{tab:p_two} (see also
Fig.~\ref{fig:Cmax_of_Np}). 
\begin{table}[htbp]
\setlength{\columnsep}{2.0cm}
  \centering
\begin{tabular}{|c|c|@{\extracolsep{1em}}lcl|@{\extracolsep{2em}}l|}
\hline
$N$&$p$&\multicolumn{3}{c|}{$C^{\text{max}}(N,p)$}&\hfill Coefficients\hfill\mbox{}\\\hline\hline
\ \ any\ \ &\ \ \ \ 1\ \ \ \ \ &$2/N$&\ \ &&$a_1=1$\\\hline\hline
2&2&0&&&$a_1=1$\\
3&2&$2/3$&&$\approx0.667\ \ $\ &$a_1=1$\\
4&2&$1/2$&&$=0.5$& $a_2=1$\\
5&2&&&$\approx0.468$&$a_1=0.298$,\ \ $a_2=0.955$\\
6&2& $\sqrt{2}/3$&&$\approx0.471$&
$a_2=a_3=\sqrt{1/2}$\\
7&2&$(1+\sqrt5)/7$&&$\approx0.462$& $a_2=\sqrt{2/(5+\sqrt5)}$,
                 \ \ $a_3=(1+\sqrt5)/\sqrt{2(5+\sqrt5)}$\\
8&2&$\sqrt3/4$&&$\approx0.433$&  $a_2=1/\sqrt6$,
                       \ \ $a_3=1/\sqrt2$,\ \ $a_4=1/\sqrt3$\\
9&2&$4\cos(\pi/7)/9$&&$\approx0.400$& $a_2\approx
0.328$,\\
&&&&&$a_3\approx0.591$, $a_4\approx0.737$\\
10&2&$\sqrt{2+\sqrt2}/5$&&$\approx0.370$& $a_2=(1-1/\sqrt2)
\sqrt{2+\sqrt2}/2$,\\
&&&&&$a_3=1/2$, \ \ $a_4=\sqrt{(2+\sqrt2)/8}$, \ \ $a_5=1/2$\\
11&2&$4\cos(\pi/9)/11$&&$\approx0.341$& $a_2\approx0.228$,
\ \ $a_3\approx0.429$,\\
&&&&&$a_4\approx0.577$, \ \ $a_5\approx0.657$\\
12&2&$\sqrt{(5+\sqrt5)/2}/6$&&$\approx0.317$& $a_2=\sqrt2/(5+\sqrt5)$, \ \ $a_3=1/\sqrt{5+\sqrt5}$,\\
&&&&&$a_4=(3+\sqrt5)/(\sqrt2(5+\sqrt5))$,\\
&&&&&$a_5=(1+\sqrt5)/(2\sqrt{5+\sqrt5})$,\\
&&&&&$a_6=(1+\sqrt5)/(5+\sqrt5)$\\\hline
\end{tabular}   
  \caption{
    Maximal nearest-neighbor concurrence and coefficients for the
optimal state, 
with $p=2$. Only non-vanishing coefficients are given explicitly.
  }
  \label{tab:p_two}
\end{table}

\subsection{Maximal nearest neighbor concurrence for $p=\frac{N}{2}$}
\label{sec:p_eq_Nhalf}

The case of half-filling, $p=N/2$, is special because the non-linear form
of the concurrence~(\ref{concurrence}) then is reduced to
\be
C(N,p=N/2)=2\max\{|z|-|v|,0\},
\label{concn2}
\ee since here the matrix elements $v$ and $y$ are equal. As mentioned
above, we may restrict ourselves to real and positive $z$, i.e.
$2|z|=z+z^*$.   When this replacement and the equality
$|v|=v$ is inserted 
in (\ref{concn2}), the concurrence can be written as the expectation
value of a Hamiltonian. We start by expressing
$z=\bra{01}\varrho\ket{10}$ and $v=\bra{00}\varrho\ket{00}$ in terms
of Pauli matrices, 
\ba
z&=&\frac14\tr\left[\varrho(\sigma^+\otimes\sigma^-)\right]\ ,\\
v&=&\frac14\tr\left[\varrho(\sigma^z-\one)\otimes(\sigma^z-\one)\right]\ 
.  
\ea 
Here $\sigma^\pm=(\sigma^x\pm i\sigma^y)$ denote the raising
and lowering operators. Due to translational invariance, the
matrix elements $z$ and $v$ are identical for {\em any} two
neighboring sites.  Thus, for non-zero concurrence and real $z$ we
arrive at 
\be C(N,p=N/2)= -\tr(\varrho H)\ , 
\ee 
where the
corresponding Hamiltonian $H$ is given by 
\be
H=\frac{1}{2N}\sum\limits_{i=1}^N\left[-\sigma_i^x\sigma_{i+1}^x-\sigma_1^y
\sigma_{i+1}^y+\sigma_i^z\sigma_{i+1}^z
+\one\right].
\label{eq:xxz_hamiltonian}
\ee 
Here we have used that the magnetization per site is
$\langle\sigma_i^z\rangle=0$ in this special case of half-filling.
The Hamiltonian~(\ref{eq:xxz_hamiltonian}) describes the Heisenberg
model with nearest-neighbor interaction in an external magnetic field.
(The relative sign between the $z$-coupling and the coupling in $x$-
and $y$-direction does not change the physical properties of the
system, as it can be absorbed by rotating the coordinate system of
every second site by $\pi$ around the $z$-axis.)  Note that the
identity in $H$ corresponds to an overall energy shift of 1/2.
The ground state energy for the Heisenberg Hamiltonian in the limit
$N\to\infty$ is well-known \cite{xxz-model} and leads to
\be
C(N\rightarrow \infty ,p=N/2)= 2 \ln2-1 = 0.386
\ee
As explained below, even for the case 
$N=20$ the concurrence is already very close to this value.

\section{Is the solution without neighboring up-spins
a local maximum?}
\label{stability}

The analytical solution for the maximal concurrence for any $N$ and
$p$ was found in \cite{wootters} by making the further restriction
(ii) of no neighboring up-spins. This leads to $y=0$ in
Eq.~(\ref{concurrence}), and one only needs to consider $z$.  The
matrix element $z$ can be written as a quadratic form in the
coefficients of Eq.~(\ref{eq:genstate}):  $z=\mya\T Z\mya$,
where $\mya$ denotes the vector with entries $a_\mu$, and $Z$ is a
matrix, the dimension of which depends on $N$ and $p$.  Thus, the
problem of maximizing the concurrence corresponds to finding the
maximal eigenvalue of $Z$, i.e.  
\be
C^{\text{max}}(N,p)=2\max\limits_k\{\lambda_k(Z)\}, 
\ee 
where
$\lambda_k(A)$ denotes the $k$th eigenvalue of $A$.  This
diagonalization problem can be solved by the Jordan-Wigner
transformation~\cite{jordan}.

It is in general a very hard problem to prove whether the
solution in~\cite{wootters} --- we will call it OW solution from now on ---
is indeed optimal. In Sec.~\ref{pequal2} we showed that for
$p=2$ and $N\ge6$ no better solution can be found, but for general
$N$ and $p$ even numerical approaches are very demanding due to the many
free
parameters and the non-linear character of the problem. As we explain
in this section, it is, however, much easier to numerically verify
whether the solution in \cite{wootters} is at least a \emph{local} maximum of
the concurrence when the restriction of no neighboring
up-spins is removed.

What we basically need to do, is to expand Eq.~(\ref{concurrence})
around the solution in \cite{wootters}. In principle, the restriction
(ii) leads to $y=0$ and thus to $\sqrt{vy}=0$, which means that $C$ as
such is not a differentiable function of the coefficients of the
state.  Thus, one has to be slightly more subtle than simply finding
the gradient of $C$.  First of all, we observe that there is a natural
grouping of the coefficients according to how many up-up pairs the
corresponding state contains: the maximal number is $p-1$, the minimal
0. A useful notation is thus to split the index on the coefficients
$a_\mu$ in a ``group'' index $\pi$ counting the number of up-up pairs
and a second index $j$ distinguishing the members of each group:
\begin{equation}
  \label{eq:double_index_def}
   a_\mu=a_{\pi j}=(\mya_{\pi})_j.  
\end{equation}
The last notation is meant to emphasize that each $\mya_\pi$ can be viewed as
a vector in its own right. The constraint (ii) allows only the
coefficients in the 0-group to be non-vanishing, i.e.\ leads to
$\mya_\pi=(0,0,\ldots,0)^\text{T}$ for $\pi>0$.

If we restrict ourselves to cases where $\text{gcd}(p,N)=1$, all
normalization factors $\mathcal{N_\mu}$ in Eq.~(\ref{mu}) are
equal to  $1/\sqrt{N}$,
and a given coefficient's contribution to $y$ only depends on its
group: 
\begin{equation}
  \label{eq:y_from_group}
  y(\mya)=\frac{1}{N}\sum_\pi \pi |\mya_{\pi}|^2.
\end{equation}
Since the number of up-up pairs gives directly the
number of down-down pairs, a similar relation holds for
$v$:
\be
  \label{eq:v_from_group}
  v(\mya)=\frac{1}{N}\sum_\pi (N+\pi-2p) |\mya_{\pi}|^2=\frac{N-2p}{N}+y.
\ee
As for $z$, things are naturally more complicated: $z$ represents the
mean value of $\ket{10}\bra{01}$ and this flip of a
down-up pair can at most create (or destroy) one
up-up pair, i.e.\ there are cross-terms between $\pi$ and
$\pi\pm 1$. In a natural matrix notation we get:
\begin{align}
  \label{eq:z_semi-block-form}
  z(\mya) & = \mya^T Z \mya
  \\
    & =
  \begin{bmatrix}
    \mya_0^T & \mya_1^T & \ldots & \mya_{p-1}^T\\
  \end{bmatrix}    
  \begin{bmatrix}
    Z_{00} & Z_{01} & 0      &           &            &  \\
    Z_{10} & Z_{11} & Z_{12} & 0         & \ldots           &  \\
     0     & Z_{21} & Z_{22} &           &            &  \\
           &   0    &        &  \ddots   &            &  \\
           & \vdots &        &           &            & Z_{p-2\:p-1}\\
           &        &        &           &Z_{p-1\;p-2}& Z_{p-1\; p-1}
  \end{bmatrix}
  \begin{bmatrix}
    \mya_0 \\ \mya_1 \\  \\ \vdots \\  \\ \mya_{p-1} 
  \end{bmatrix}
  ,
\end{align}
where the $Z_{\pi\pi'}$ are still matrices with indices running
inside the groups. If we take the derivative of $z$ at OW's
solution $\mya=\myb$, we get
\begin{align}
  \label{eq:grad_z}
  \frac{\partial z}{\partial a_{\pi j}} (\myb)
  &= 2 [Z\myb]_{\pi j} 
  \nonumber \\
  &= 2 \delta_{\pi 0} \; [Z_{00} \myb_{0} ]_j
  +
   2 \delta_{\pi 1} \; [ Z_{10}\myb_{0}]_j
  \nonumber \\
  &=  2 \delta_{\pi 0} \;  \;z(\myb)\; [\myb_{0}]_j
  +
  2 \delta_{\pi 1} \;  [ Z_{10}\myb_{0}]_j
  ,
\end{align}
where have used that $\myb_\text{0}$ is an eigenvector (belonging
to the largest eigenvalue) of $Z_{00}$. Since we have the further
constraint $|\mya|^2=1$, only the component of the gradient
perpendicular to the OW solution is interesting, i.e.\ only the
second term in Eq.~(\ref{eq:grad_z}) needs to be considered. From the
$\delta_{\pi 1}$ we see that any local first order improvement of
OW's solution can only be found by adding a small component from the
space with exactly one up-up pair. Of course, when we do
that $y$ will no longer be zero, the square root in
Eq.~(\ref{concurrence}) will grow and this will possibly cancel
the benefits we could get from $z$ alone. A priori we should test all
directions in the $\pi=1$ space, but since $y$ does not discriminate
between them, the one suggested by $z$ suffices: if $C$ does not grow
when adding an infinitesimal component along $Z_{10}
\myb_{0}$, the solution of OW is a local maximum. The exact
condition for this to happen reads:
\be
  \label{eq:loc_max_ineq}
  2 |Z_{10}\myb_{0}| 
  < \sqrt{v(\myb)}
  =\sqrt{1-2\frac{p}{N}}.
\ee
If, on the other hand, this inequality is fulfilled in the other
direction, we know that OW's solution can be improved even locally
and is thus for sure not the global maximum.

In Fig.~(\ref{complete}) we will present the results of a numerical
investigation of Eq.~(\ref{eq:loc_max_ineq}) for various $N$ and $p$,
together with the results of the non-perturbative investigation
described in the following section.

\section{Mapping to a series of Hamiltonian problems}
\label{sec:map}

In Sec.~\ref{sec:p_eq_Nhalf} we saw how the non-linear problem of
optimizing the concurrence could be stated as a linear one in the
special case of $p=N/2$. A method which exploits this idea further
has recently been introduced by Wolf {\it et al.}~\cite{cirac}: The
maximal value of Eq.~(\ref{concurrence}) on translationally invariant
states (for non-zero concurrence and $z$ real
and positive) coincides with the
maximum of
\begin{equation}
C(s)= z + z^* 
         - \left(sy+\frac{1}{s}v\right)\ ,
\label{eq:ciractrick}
\end{equation}
where the maximization is done over all translationally invariant
states \emph{and} all $s > 0$. To see that this holds, simply note
that
the minimal value of $sy+v/s$ with respect to $s$ is $2\sqrt{yv}$,
which is reached for $s=\sqrt{v/y}$.
Thus, after performing first the maximization of
Eq.~(\ref{eq:ciractrick}) with respect to $s$,
one returns to the original problem. 

If on the other hand we optimize in the opposite order, then for a given 
$s$ we can generalize the method from Sec.~\ref{sec:p_eq_Nhalf} and 
relate $C(s)$ to the expectation value of a Hamiltonian. Explicitly we
have
\begin{equation} 
C(s) = -\tr[\varrho H(s)], 
\end{equation}
where 
\begin{equation}
H(s)=\frac{1}{2N}\sum\limits_{i=1}^{N}\left[-\sigma_i^x\sigma_{i+1}^x-
  \sigma_i^y\sigma_{i+1}^y+\Delta(s)\left(\sigma_i^z\sigma_{i+1}^z+
    \one\right)+B(s)\sigma_i^z\right].
\label{eq:s_hamiltonian}
\end{equation}
This Hamiltonian corresponds to the quantum XXZ-model with an external
magnetic field $B(s):=s-1/s$ and an anisotropy parameter
$\Delta(s):=(s+1/s)/2$. 
For $s=1$ Eq. (\ref{eq:s_hamiltonian})
 reduces to Eq. (\ref{eq:xxz_hamiltonian}).
The XXZ-model is completely integrable, and in
principle all eigenvalues and eigenstates of the
Hamiltonian~(\ref{eq:s_hamiltonian}) can be found using the Bethe
ansatz~\cite{gaudin}. Unfortunately for general $p$, corresponding to 
general magnetization, the solution is very difficult to handle
even in the limit $N\rightarrow\infty$~\cite{xxz-model}, and we
have to resort to numerical solutions.

\begin{figure}[ht]
\includegraphics*[width=9.5cm]{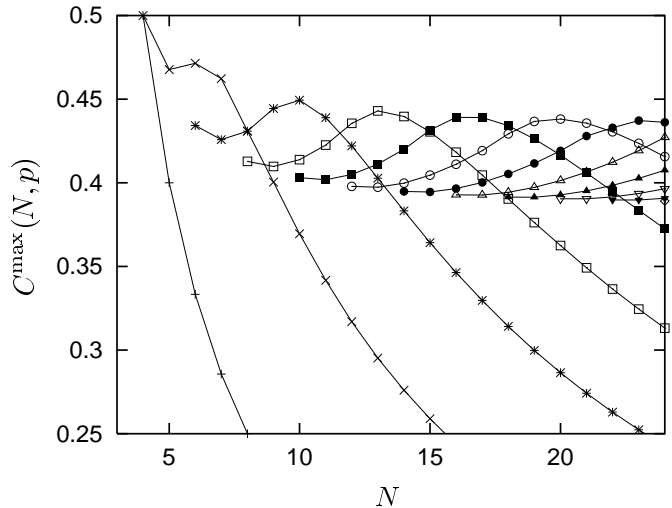}
\caption[]{Maximal 
  nearest-neighbor concurrence for $N=4,\ldots,24$. Each line and
  symbol type corresponds to a fixed $p$: 
  the left-most curve illustrates the case $p=1$ and is marked by the
  symbol ``+''. The second curve from the left shows $p=2$,
and so forth. 
The explicit legend is also given in 
  Fig.~\ref{plots}. Curves with higher $p$'s peak at higher $N$, the
  optimal value scaling roughly as $3p$.}
\label{fig:Cmax_of_Np}
\end{figure}

For a given $s$, we minimize Eq. (\ref{eq:s_hamiltonian}) by 
direct diagonalization. It turns out that the dependence
of the ground-state energy on $s$ is rather simple, such that the final 
optimization with respect to $s$ provides only a modest complication.
In Fig.~\ref{fig:Cmax_of_Np} we plot
the results obtained for all relevant $p$ as $N$ ranges from $1$ to $24$.
``Relevant'' $p$ are those with $p\leq N/2$, as the concurrence is
identical for occupation number $p$ and $N-p$.
For a fixed $p$ the maximal concurrence has in general the following
behavior as a function of $N$: from the first relevant $N=2p$, it
slightly decreases and then 
grows to a maximum around $N=3p$, after which it tends slowly to $0$ as
$N\rightarrow\infty$. 

\begin{figure}[htp]
\includegraphics*[width=9.5cm]{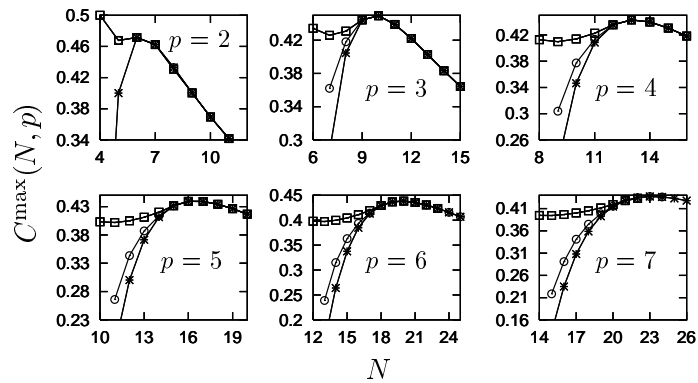}
\caption[]{Maximal nearest-neighbor
concurrence for $p=2,\ldots,7$ as function
  of the number of sites: OW result (crosses), 
  perturbation theory 
(circles),
  and exact solution (squares). Note that the perturbation theory and
  the exact solution in all cases meet the OW result at the same
  $N$. This implies that OW's solution is never a local maximum
  without being also the global maximum.}
\label{complete}
\end{figure}

In order to analyze more carefully when OW's result is optimal, in
Fig.~\ref{complete} we show parts of the $C^\text{max}(N)$ curves for
the OW constraint of no neighboring up-spins, for the improvement via
perturbation theory, and for the exact solution. As can be clearly
seen, optimal states with no neighboring up-spins can be found
for small numbers of sites, namely for $N$ smaller than the value
for which  the concurrence 
has its maximum for a given  $p$.
Note also that the local optimization via perturbation theory and
the exact calculation always start to
deviate from the OW curve at
the same $N$. This implies that when the OW state is locally optimal,
it is  also globally optimal.

\begin{figure}[htp]
\includegraphics*[width=9.5cm]{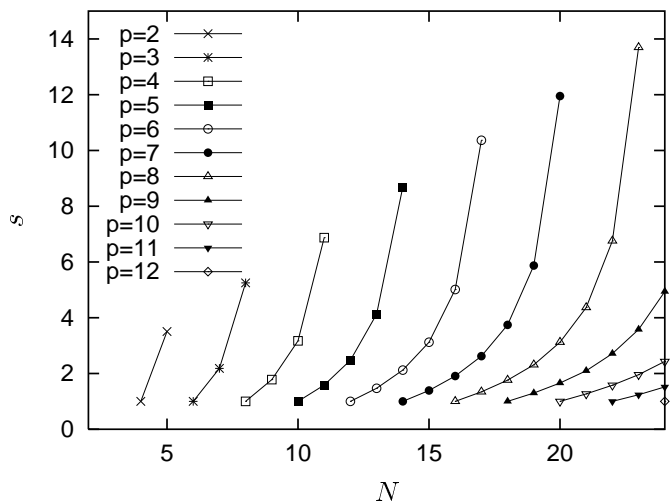}
\caption[]{\small Optimal value for the parameter $s$ as a function of
$N$ and $p$. 
}
\label{plots}
\end{figure}

It is furthermore interesting to study the values for the optimal 
$s$ as a function of
$N$ and $p$, see Fig.~\ref{plots}. 
These values immediately give the physical interaction that
leads to the ground state with maximal nearest-neighbor
concurrence. Notice that 
$s=1$ for $p=N/2$, and that for any value of $p$ the optimal value
of $s$ goes to infinity with $N\rightarrow\infty$. This indicates
that the OW solution is indeed optimal for large $N$, and that the
occurrence of neighboring up-spins for the maximization of the
concurrence is a true finite size effect.

Let us also mention the exceptional case $N=8$. In \cite{wootters},
no solution with a concurrence above the asymptotic value
$C^\text{max}(N\rightarrow\infty)\approx 0.434$ 
was found for any $p$. 
Is this an artifact of the restriction to no neighboring up-spins,
or a true finite size effect? Our studies show that the latter is
true: we find
$C^\text{max}(N=8,p=3)\approx 0.431$. Thus,
together with the results presented in Table \ref{tab:p_two}, 
the occupation number $p=2$ is shown
to be optimal for
$N=8$. However, 
$C^\text{max}(N=8,p=2)=\sqrt{3}/4\approx 0.433<C^\text{max}
(N\rightarrow\infty)$
holds --- a peculiar feature of a spin chain with $8$ sites.

\section{Beyond nearest neighbor entanglement}
\label{beyondnearest}

A natural extension to the analysis done in the previous section is to
examine entanglement between spins separated by at least one site. In
principle, the formalism of Sec.~\ref{pequal2} can be applied
to the maximization problem of the concurrence between spins separated by
any number of sites. This approach yields results similar to those of
Sec.~\ref{pequal2}. For example, when maximizing the next-nearest neighbor
entanglement for $p=2$, we find that using the constraint ``no next-nearest
neighbors are both up'', i.e. taking $a_2=0$, does not
decrease the concurrence for $N\ge8$.

In this section we will treat the issue of reducing the
$q$th-neighbor maximization problem of a ring with $N$ sites
to a nearest-neighbor maximization problem of a ring with less than
$N$ sites. In contrast to the previous sections, we will not fix the
number of up-spins; we simply look for the maximum of $C$ with
translational invariance and no superpositions of different $p$.

Assume first that $q$ divides $N$, 
i.e. $q|N$. 
From now on we will denote the maximal $q$th-neighbor 
concurrence of a ring with $N$ sites by $C^{\text{max}}(N,q)$. 
 We denote by $\ket N^{\text{opt}}$ a state
that reaches this value and by $p^\text{opt}$ the corresponding number
of up spins. 

\begin{figure}[ht]
\includegraphics*[width=4.6cm]{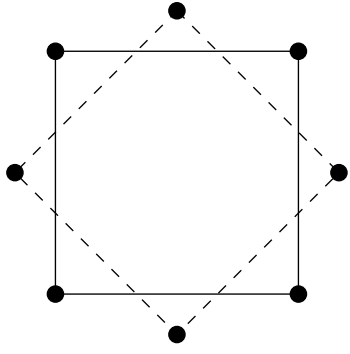}
\caption[]{Decomposition of the ring into smaller rings, for the case $N=8$
and next-nearest neighbors.
}
\label{ringbeyond}
\end{figure}

Consider now the decomposition of the large ring into $q$ smaller
rings. Each of them consists of every $q$th qubit of the large 
ring and has  $N/q$ sites.
The example of a ring with 8 sites and next-nearest-neighbor
entanglement, i.e. the case $N=8$ and $q=2$, is illustrated in
Fig.~\ref{ringbeyond}.
The state that maximizes the nearest-neighbor
concurrence of a small ring, $C^{\text{max}}(N/q,1)$, will be denoted
by $\ket{N/q}^{\text{opt}}$.  The $q$-fold tensor product of this
state is a candidate for the state $\ket N^{\text{opt}}$: it is
translationally invariant since $\mathcal{T}$ can be decomposed into a
permutation of the set of small rings and translations within them. We
thus have:
\begin{equation}
C(\underbrace{\ket{N/q}^{\text{opt}}\otimes\cdots\otimes\ket{N/q}^{\text{opt}}}_
{q\,\text{times}},q)\le C^{\text{max}}(N,q)=C(\ket N^{\text{opt}},q)
\label{eq:cnnlb}
\end{equation}
To show than one cannot do better than using this candidate, we
note that the entanglement between the first and the $q$th site of
the state $\ket N^{\text{opt}}$ is calculated by tracing over all
states but the first and $q$th. In particular, we trace over all but
one small ring. Let us denote the state obtained by tracing out the
$q-1$ irrelevant rings by $\varrho^\text{(s.r.)}$, i.e.  
\begin{equation}
\varrho^{(\text{s.r.})}=
\tr_{\text{other rings}}\ket N\bra N^{\text{opt}}\ .  
\end{equation}
Therefore, 
\begin{equation}C(\ket
  N^{\text{opt}},q)=C(\varrho^{(\text{s.r.})},1).  
\end{equation}
Now, because the full $\varrho$ was translationally invariant,
$\varrho^{(\text{s.r.})}$ is also translationally invariant. This means
by definition that
\begin{equation}
  \label{eq:cgen_less_copt}
  C(\varrho^{(\text{s.r.})},1) \le C^\text{max} (N/q,1) 
  = C(\ket{N/q}^\text{opt},1) .
\end{equation}
Since the tensor product of Eq.~\eqref{eq:cnnlb} saturates this
inequality, it must be optimal. We conclude that when $q\;|N$,
\begin{equation}
  \label{eq:q_div_N_cmax}
  C^\text{max}(N,q)=C^\text{max}(N/q,1)
  .
\end{equation}

Let us now drop our assumption $q|N$ and assume
 that $N$ and $q$ are coprime, $\mygcd{N}{q}=1$. If one then tries
to construct ``small rings'' by taking steps of $q$ sites, one will
eventually visit all sites, i.e. one only finds one such ring. This simply
reflects the fact
that $\mathcal{T}^q$ generates the same group of translations
as $\mathcal{T}$ does. Then it is clear that by reordering of the
spins one can construct a ring on which all spins that were formerly
separated by $q$ sites are now adjacent \emph{and} on which
translational invariance is equivalent to translational invariance
with respect to the original ordering. Therefore, when $\mygcd{N}{q}=1$, 
\begin{equation}
C^{\text{max}}(N,q)=C^{\text{max}}(N,1).  
\end{equation}

In general, constructing small rings by taking steps of $q$ sites will
lead to $\mygcd{N}{q}$ rings, each of length $\tilde{N}=N\;/\mygcd{N}{q}$.
On these rings the former $q$th  neighbors are now nearest
neighbors, and an optimal
state $\ket{N,q}^\text{opt}$ can be constructed similarly to the
$q\;|N$ case above as a tensor product of small ring states
$\ket{\tilde{N},1}^\text{opt}$.
In this way, the problem can always be reduced to the case of nearest
neighbors with the result that for any $N$ and $q$:
\begin{equation}
  \label{eq:gen_qth_nn}
  C^\text{max}(N,q)=C^\text{max}(N/\mygcd{N}{q},1)
  .
\end{equation}

The method of this section naturally generalizes to other symmetry
groups acting on $N$ qubits: If we choose to optimize a given pair's
entanglement, the symmetry group will give some restrictions, but
these may ``factor'' so that tensor product solutions of smaller
problems can be applied. In the ring symmetry case,
Eq.~(\ref{eq:gen_qth_nn}) shows that the reduction is maximal in the
sense that there is essentially only one class of problems, namely nearest
neighbors on a ring with translational symmetry. This is 
due to the particularly simple structure of the group of
translations; in general one symmetry group may give rise to many
fundamentally different optimization problems.

\section{Discussion and Summary}
\label{sec:discuss}

In this paper we have studied maximal bipartite
entanglement in translationally invariant spin chains with
periodic boundary conditions, where our only restriction was to
consider eigenstates of the $z$ component of the total spin.
We have concentrated on bipartite entanglement, as 
entanglement in this case is
fully understood, contrary to multipartite entanglement.

Naturally, some related
questions arise:
\begin{itemize}
\item What is the maximal entanglement between one spin and the rest of the
ring?
\item What is the possible structure of three- or
multi-party entanglement in the 
ring?
\item What is the optimal nearest-neighbor entanglement on a 2-dimensional
lattice?
\end{itemize}
Here we will only  answer the first question: 
states with fixed occupation number $p$ are the eigenstates
of the $z$ component of the total spin, i.e. they are invariant
under a global rotation around the $z$-axis,
$U=\bigotimes_{i=1}^{N}\sigma_i^z$. This leads to
$[\sigma^z,\varrho^{(1)}]=0$, where
$\varrho^{(1)}$ denotes the one-particle reduced density matrix,
which is identical for all sites
due to translational invariance. Thus, $\varrho^{(1)}$
is diagonal, i.e. $\varrho^{(1)}=(\one +s^z\sigma^z)/2$.
Here $s^z=\langle\sigma^z\rangle$ is  the magnetization 
in $z$-direction for one site. The explicit form of the reduced density matrix
is therefore
\be
\label{eq:rho1}
\varrho^{(1)}=\frac{1}{N}\mat{(N-p) & 0\\ 0 & p }\ . 
\ee
Note that this form is independent of the choice of coefficients
$a_\mu$, and results only from translational invariance and the
condition of 
fixed $p$.
The von Neumann entropy of this density matrix is a good measure
for the entanglement of one qubit with the rest of the chain.
Therefore, we find maximal entanglement 
in this sense when $p=N/2$.

In summary, we have investigated the
maximal nearest-neighbor concurrence in 
translationally invariant entangled rings,
where the number of up-spins $p$ is fixed. We presented analytical results
for small $N$ and $p=2$. 
It also turned out that the solutions in \cite{wootters} can be
improved by simple perturbation theory. We then 
used a linearized version of the concurrence, which
simplified the numerical calculations, and with this method determined the  
maximal entanglement for any $p$ with $N\le 24$. 
The maximal entanglement was shown to be related to the ground state
energy of an XXZ Hamiltonian. Let us emphasize our observation that a
state with maximal nearest-neighbor entanglement corresponds to a
Hamiltonian with only nearest-neighbor interactions.
For small numbers of sites $N$ and fixed $p$ we
 gave  the explicit structure of the  Hamiltonian 
which leads to  the state with maximal 
entanglement of nearest neighbors.
We found the peculiar result that for the exceptional case $N=8$
the maximal entanglement is below the entanglement in the
thermodynamic limit. Finally, we reduced the problem of finding
the maximal entanglement between spins 
which are separated by more than one
 site to the problem of nearest-neighbor entanglement.
 
 This work has been supported by the Deutsche Forschungsgemeinschaft
 (SFB 407 and Schwerpunkt ``Quanteninformationsverarbeitung"), the
 European Union RTN ``Cold Quantum Gases'', and the Danish Natural Science
 Research Council.
 

\appendix*

\section{}
\label{sec:lagrange}

Here we describe a method to find the maximal concurrence
and the according state for the case $N\geq 8$ and $p=2$.
We take the normalization constraint, namely $\sum_ia_i^2=1$,
into account via Lagrange multipliers.

 For {\em even} $N$ we then have to solve the system of equations
\bea
a_3+2a_2\lambda &=& 0\ , \nonumber \\
a_2+a_4+2a_3\lambda &=& 0\ ,  \nonumber \\
a_3+a_5+2a_4\lambda &=& 0\ ,  \nonumber \\
&...&\  \nonumber \\
a_{N/2-2}+\sqrt{2}a_{N/2}+2a_{N/2-1}\lambda &=& 0\ , \nonumber \\
\ \ \ \ \sqrt{2}a_{N/2-1}+2a_{N/2}\lambda &=& 0\ .
\label{syseven}
\eea
where $\lambda$ is the Lagrange multiplier.
For {\em odd} $N$ the last two equations in the system of equations
 look slightly different:
\bea
+a_3+2a_2\lambda &=& 0\ ,  \nonumber \\
a_2+a_4+2a_3\lambda &=& 0\ ,  \nonumber \\
a_3+a_5+2a_4\lambda &=& 0\ ,  \nonumber \\
&...&\  \nonumber \\
a_{(N-5)/2}+a_{(N-1)/2}+2a_{(N-3)/2}\lambda &=& 0\ , \nonumber \\
a_{(N-3)/2}+a_{(N-1)/2}+2a_{(N-1)/2}\lambda &=& 0\ .
\label{sysodd}
\eea
In both cases we now multiply the first equation with $a_2$, the second
with $a_3$ and so on, and then add all equations.
This leads to the expression
\begin{equation}
C_{max}=-\frac{4\lambda}{N}
\label{conc}
\end{equation}
for the maximal concurrence. Note that this equality holds for
both even and odd $N$. By inserting the first  equation
in the system of equations~(\ref{syseven})
into the second, the new one into the third, and consecutively
until the last equation, one finds a polynomial 
equation for $\lambda$
of order $N/2-1$ for even $N$.
By proceeding with (\ref{sysodd}) in the same way one arrives at
a polynomial equation
of order $(N-3)/2$ for odd $N$.
For small $N$, these polynomial equations can be solved analytically, and
for larger $N$ one can still find the zeros of the polynomial 
numerically.

Let us illustrate this method with a simple explicit example, 
namely the case $N=8$ and $p=2$. This is also a particularly
interesting case, because in~\cite{wootters} in was shown that the
maximal concurrence for this case (with the constraint of no
two neighboring spins being up) is {\em smaller} than 
for the thermodynamical limit $N\rightarrow\infty$. 

The system of equations for $N=8$ and $p=2$ is
\bea
a_3+2a_2\lambda &=& 0\ ,  \nonumber \\
a_2+\sqrt{2}a_4+2a_3\lambda &=& 0\ ,  \nonumber \\
\sqrt{2}a_3+2a_4\lambda &=& 0 \ .
\eea
 By inserting these 
equations successively into each other we arrive at 
\begin{equation}
4 \lambda^3-3\lambda =0 \ .
\end{equation}
In order to maximize the concurrence in Eq.~(\ref{conc})
we have to find the minimal solution for $\lambda$, which
is $\lambda = -\sqrt{3}/2$. Therefore $C(N=8,p=2)=\sqrt{3}/4$.
The coefficients of the optimal entangled  state are then easily found
to be $a_2=\sqrt{1/6}, a_3=\sqrt{1/2}$ and $a_4=\sqrt{1/3}$.
The results of other cases of small $N$ are given in the main text.


\end{document}